\documentclass[12pt]{article}
\usepackage{amsmath,amsfonts,amssymb,mathrsfs,cite}
\usepackage{color,graphicx,shortvrb,epsfig}
\usepackage{latexsym}
\usepackage[a4paper]{geometry}
\usepackage[in]{fullpage}
\newlength{\myem}
\newcounter{mysubequation}[equation]

\makeatletter
\renewcommand{\section}{\@startsection{section}{1}{0em}{-\baselineskip}%
{\baselineskip}{\normalfont\large\bfseries}}
\renewcommand{\subsection}%
{\@startsection{subsection}{2}{0em}{-0.7\baselineskip}%
{0.7\baselineskip}{\normalfont\bfseries}}
\makeatother

\newcommand{\ba}{\begin{array}}
\newcommand{\ea}{\end{array}}
\newcommand{\bd}{\begin{displaymath}}
\newcommand{\ed}{\end{displaymath}}
\newcommand{\bi}{\begin{itemize}}
\newcommand{\ei}{\end{itemize}}
\newcommand{\benu}{\begin{enumerate}}
\newcommand{\eenu}{\end{enumerate}}
\newcommand{\be}{\begin{equation}}
\newcommand{\ee}{\end{equation}}
\newcommand{\bea}{\begin{eqnarray}}
\newcommand{\eea}{\end{eqnarray}}

\def\gsim{\ \raisebox{-.45ex}{\rlap{$\sim$}} \raisebox{.45ex}{$>$}\ }
\newcommand{\ie}{{\it i.e.}}

\def\pee{{{\rm P_{e e}}}}
\def\pmue{{{\rm P_{\mu e}}}}
\def\pmumu{{{\rm P_{\mu \mu}}}}
\def\pmumum{{{\rm P^m_{\mu \mu}}}}
\begin{document}
%
%
%
%
\begin{flushright}
\end{flushright}
\vspace*{1cm}
\setcounter{footnote}{-1}
    {\begin{center}
    {\LARGE{\bf{{Mass Hierarchy Determination for $\theta_{13}=0$ and Atmospheric Neutrinos}}}}
    \end{center}}
\renewcommand{\thefootnote}{\fnsymbol{footnote}}
\vspace*{1cm}
                {\begin{center}
                {{\bf
                Raj Gandhi $^{a, \,\!\!\!}$
                \footnote[1]{\makebox[0.cm]{}
                \sf raj@mri.ernet.in},
                Pomita Ghoshal $^{b, \,\!\!\!}$
                \footnote[2]{\makebox[0.cm]{}
                \sf pomita@theory.tifr.res.in},
                Srubabati Goswami $^{a,c, \,\!\!\!}$
                \footnote[3]{\makebox[0.cm]{}
                \sf sruba@prl.res.in},
                S Uma Sankar $^{d, \,\!\!}$
                \footnote[5]{\makebox[0.cm]{}
                \sf uma@phy.iitb.ac.in}
                }}
                \end{center}}
\vskip 1.2cm
{\small
                \begin{center}
                $^a$ Harish-Chandra Research Institute, Chhatnag Road, \\
                     Jhunsi, Allahabad 211 019, India\\[4mm]
                $^b$ Department of Theoretical Physics,Tata Institute of Fundamental Research, \\
                     Colaba, Mumbai 400 005, India\\[4mm]
                $^c$ Physical Research Laboratory, \\
                     Navrangpura, Ahmedabad 380 009, India\\[4mm]
                $^d$ Department of Physics, Indian Institute of Technology, 
Powai,\\
                     Mumbai 400 076, India
                \end{center}}

\vspace*{0.5cm}
\date{\today}
%
%
%
%
\begin{abstract}
We examine the possibility of determining the neutrino mass hierarchy in the limit $\theta_{13} = 0$ using
atmospheric neutrinos as the source. In this limit, in which $\theta_{13}$ driven matter effects are absent,  independent measurements of $\Delta_{31}$ and $\Delta_{32}$
can, in principle,  lead to hierarchy determination. Since the difference between these two is
$\Delta_{21}$, one needs  an experimental arrangement where $\Delta_{21} L/E \gtrsim 1$ can be achieved. 
This condition can be satisfied by
atmospheric neutrinos since they have a large range of energies and baselines. In spite of this, we find that hierarchy
determination in the $\theta_{13}=0$ limit with atmospheric neutrinos is 
not a realistic possibility, even in conjunction with an apparently synergistic beam experiment like T2K or NO$\nu$A.
We discuss the reasons for this, and also in the process clarify
the conditions that must  be satisfied in general 
for hierarchy determination if $\theta_{13} = 0$.
\end{abstract}
%
%
%
{\bf{PACS: 14.60.Pq,14.60. Lm.,13.15.+g}}

\newpage
\renewcommand{\thefootnote}{\arabic{footnote}} 
\setcounter{footnote}{0}

\section{Introduction} 
The neutrino mass hierarchy is an important discriminator between 
large classes of models of unified theories \cite{Albright:2004kb}. 
Its determination remains one of the outstanding problems facing
neutrino physics. The term hierarchy refers to the position of the  
third mass eigenstate with respect to the other two (relatively) degenerate
states. When the  hierarchy is normal (NH), the third mass eigenvalue $(m_3)$ 
is the largest and
hence $\Delta_{31} = m_3^2 - m_1^2$ and $\Delta_{32} = m_3^2 -
m_2^2$ are both positive. On the other hand, when it is  inverted (IH), $m_3$
is the smallest, leading to $\Delta_{31}$ and $\Delta_{32}$ 
being negative. 

The most widely discussed method for hierarchy determination relies
on sizable matter effects at long baselines.
The baselines of the foreseeable super-beam experiments are $1000$
Km or less. Therefore in these experiments appreciable matter
effects do not develop, although it may be possible to have some
hierarchy sensitivity in the $\nu_\mu \rightarrow \nu_e$ channel
\cite{Narayan:1999ck,Mena:2006uw} provided $\theta_{13}$ is large
enough (\ie  $\sin^2 2 \theta_{13} \geq 0.04$). However, 
this channel is seriously compromised  by the
($\delta_{CP}$,hierarchy) degeneracy
\cite{Minakata:2001qm,Barger:2001yr}. Thus hierarchy determination
via this 
channel requires data from a number of
complementary experiments. The degeneracy problem is reduced in the
case of a detector located at the magic baseline of $\sim$ 7000 km
\cite{Barger:2001yr,Huber:2003ak,Smirnov:2006sm}. However, this requires high luminosity sources
such as neutrino factories or beta-beams
\cite{Bandyopadhyay:2007kx}, likely to be
available only in the  far future.
Recently synergy in T-conjugated channels were explored to 
determine hierarchy for shorter baselines with small matter effects
for $\sin^2 2\theta_{13} \gsim 0.03$
 \cite{Schwetz:2007py}. 

Independent of the value of $\theta_{13}$, $\Delta_{31}$
is the largest mass squared difference for NH, whereas
for IH, $\Delta_{32} = \Delta_{31} - \Delta_{21}$ has the largest magnitude. In
principle, one can exploit this difference to determine the
mass hierarchy if one can measure $|\Delta_{31}|$ and $|\Delta_{32}|$ individually and precisely \cite{Learned:2006wy}. 
Detection of the interference effect in vacuum neutrino oscillations 
\cite{Petcov:2001sy,Choubey:2003qx,Nunokawa:2005nx} 
and the measurement of the phase of monoenergetic ${\bar{\nu_e}}$ \cite{Minakata:2007tn} 
have been proposed to determine the hierarchy.

With atmospheric neutrinos as the source, wide ranges in energy
$(E)$ and baseline $(L)$ become available and it is possible to
observe large resonant matter effects
\cite{Mikheyev:1989dy,Gandhi:2004md,Gandhi:2004bj,Gandhi:2007td,
Petcov:2005rv,Akhmedov:2006hb}.
For values of $\theta_{13}$ much below the CHOOZ upper bound
\cite{Apollonio:1999ae}, the distance a neutrino needs to travel
for these effects to be observable, becomes larger than the diameter
of the earth \cite{Gandhi:2004md}. Therefore, hierarchy
determination with atmospheric neutrinos using matter effects 
also requires a moderately large value of $\theta_{13}$ 
\cite{Gandhi:2007td,Petcov:2005rv} 
($\sin^2 2\theta_{13} \gsim$ 0.05).

There is a large class of models which predict $\theta_{13}$ to be exactly 
zero  
due to some symmetry principle \cite{Albright:2006cw} 
\footnote{It is of course likely that
such  symmetres are broken at low energies and 
$\theta_{13}$ would acquire a small non-zero value \cite{Joshipura:2002kj,Joshipura:2002gr,Mei:2004rn,Grimus:2004yh,Dighe:2008wn}. 
Hierarchy 
determination using matter effects
driven by very small values of $\theta_{13}$ 
entails  measuring the $\nu_e \rightarrow \nu_\mu$
probability at the magic baseline using a beta-beam 
for $\sin^2 2\theta_{13} \geq 10^{-2}$ \cite{Agarwalla:2006vf},
or with a  neutrino factory beam for 
$\sin^2 2\theta_{13} \geq 10^{-3}$ \cite{Barger:2000cp}.}. 
Recent global analysis of neutrino data indicates 
a small preference for a non-zero $\theta_{13}$, $\sin^2 2\theta_{13}
\simeq 0.05$ \cite{Fogli:2008jx,Schwetz:2008er}.  
However, at present the statistical significance of this result 
is only  at the level of 
$ \sim 1\sigma$. In addition, the latest results from MINOS show an excess of 
electron events \cite{Diwan:2009pn}. This 
may also be an indication for a non-zero 
$\theta_{13}$ \cite{Fogli:2009ce}.  


In the present work we consider the 
problem of determination of the sign of atmospheric neutrino 
mass difference in the limit $\theta_{13} \rightarrow 0$. This 
is a challenging problem  requiring high precision measurements.  
The general requirements for this were discussed in 
\cite{deGouvea:2005hk}. It was argued that one would need 
measurements of $P_{\mu \mu}$ and $P_{\bar{\mu} \bar{\mu}}$ 
at different $L/E$ with atleast one $L/E$ satisfying
$\Delta_{21} L/E \geq 1$.  In this conenction  
the utility of a broadband 
accelerator beam or atmospheric neutrinos capable of providing different 
values of $L/E$ within the same experimental set up was 
pointed out in \cite{deGouvea:2005hk}. 
Reference \cite{deGouvea:2005mi} made a detailed implementation of 
the ideas discussed in 
\cite{deGouvea:2005hk} 
in the context of neutrino factories and superbeams using two narrow band beams at different $L$ or one broad band beam. 
However, no detailed  quantitative analysis of hierarchy determination for 
$\theta_{13}=0$  using 
atmospheric neutrinos 
had been done  in the literature so far. 
In this paper, we undertake this task and 
perform the  analysis
using atmospheric neutrinos   
in conjunction with 
long baseline experiments.  
Our study shows that despite the fact that atmospheric neutrinos provide 
a large range of $L$ and $E$,
practically speaking, this is an 
impossible prospect.  
We  discuss in detail the conceptual and experimental issues that lead to
this conclusion. 


\section{$P_{\mu \mu}$ in the limit $\theta_{13} =0$} 

For $\theta_{13} = 0$, the probabilities involving the 
electron flavour depend only on $\Delta_{21} = m_2^2 - m_1^2$ because the
mass eigenstate $\nu_3$ decouples from $\nu_e$. Hence
probabilities involving $\nu_e$ are not sensitive to the sign of
$\Delta_{31}$ 
and to determine the hierarchy, other
oscillation and survival modes must be explored. In the
case of atmospheric neutrinos, the $\nu_\mu \to \nu_\mu$ survival
probability provides the most effective channel, especially
from the detection viewpoint.

As explained above, for different hierarchies the
relative magnitudes of $\Delta_{31}$ and $\Delta_{32}$ are different.
The difference in  magnitudes of $\Delta_{31}$ and $\Delta_{32}$ 
for NH and IH is
proportional to
$\Delta_{21}$. To exploit the corresponding difference in $\pmumu$, 
one needs values
of $L/E$ such that $\Delta_{21} L/E \gtrsim 1$ or $L/E \gtrsim 10^4$.
Such values of $L/E$ can be obtained for $E \lesssim 1$ GeV and
$L \sim 10^4$ Km. A large flux of atmospheric
neutrinos with these baselines and energies is readily available.

For such long distances,
the neutrinos propagating through earth's matter experience
a potential $V = \sqrt{2} G_F n_e$  ($G_F$ is the Fermi Constant
and $n_e$ is the ambient electron number density). This gives rise
to an effective mass squared matter term, $A = 2 E V$ 
which can be expressed as
\be
A =  2 \times 7.6 \times 10^{-5}  ~\rho {\mathrm{(g/cm^3)}} ~Y_e
~E {\mathrm{(GeV)}} ~[{\mathrm{eV^2}}]
\label{matpot}
\ee
where $\rho$ is the
average density of matter along the neutrino path and
$Y_e = 0.5$ is the electron
fraction per nucleon.

Including matter effects,
the expression for ${\mathrm{P_{\mu \mu}}}$ for $\theta_{13}=0$ is
\bea
{\rm{P^m_{\mu \mu}}} & = & 1 - 4 \cos^4 \theta_{23} \sin^2
\theta_{12}^m \cos^2 \theta_{12}^m
 \sin^2 \left( \Delta_{21}^m \frac{L}{4E} \right)
\nonumber \\
& & - 4 \cos^2 \theta_{23} \sin^2 \theta_{23} \sin^2
\theta_{12}^m \sin^2 \left( \Delta_{31}^m \frac{L}{4E} \right)
\nonumber \\
& & - 4 \cos^2 \theta_{23} \sin^2 \theta_{23} \cos^2 \theta_{12}^m
\sin^2 \left( \Delta_{32}^m \frac{L}{4E} \right) \label{eqpmumu0}
\eea
where, the squared mass-differences in matter are
\begin{eqnarray}
\Delta_{21}^m & = &  m_{2m}^2 - m_{1m}^2
\nonumber \\
\Delta_{31}^m & = &  m_{3m}^2 - m_{1m}^2 = \Delta_{31} - \frac{1}{2}
\left[ \Delta_{21} + A - \Delta_{21}^m \right]
\nonumber \\
\Delta_{32}^m & = &  m_{3m}^2 - m_{2m}^2 = \Delta_{31} - \frac{1}{2}
\left[ \Delta_{21} + A + \Delta_{21}^m \right] \label{sqmd}
\end{eqnarray}
and,
\begin{equation}
\Delta_{21}^m = \sqrt{(\Delta_{21} \cos 2 \theta_{12} - A)^2 +
(\Delta_{21} \sin 2 \theta_{12})^2} \label{del21m}
\end{equation}
\begin{equation}
\sin 2 \theta_{12}^m = \sin 2 \theta_{12}
\frac{\Delta_{21}}{\Delta_{21}^m}. \label{sin12m}
\end{equation}
From the above equations we see that the condition
$\Delta_{21} \cos 2\theta_{12} = A$ defines a resonance energy
related to the solar mass-squared difference $\Delta_{21}$,
\bea
E_s^R (GeV)  = \frac{\Delta_{21} \cos 2\theta_{12}}{0.76 \times 10^{-4}
\rho (g/cm^3)}.
\label{esolres}
\eea
The present best fit value of $\Delta_{21} \sim 8 \times 10^{-5}$ eV$^2$
\cite{:2008ee,Bandyopadhyay:2008va,Maltoni:2004ei} gives
$E_s^R \approx 0.06-0.2$ GeV.

In the third column of 
Table 1 we list the values of $E_s^R$ for the various baselines. 
Note that for longer baselines, the average density is larger because the neutrinos
pass through the inner mantle ($5000 \le L \le 10000$ Km) and the core ($L \ge 10000$ Km).

\begin{table}[htb]
\begin{center}
\begin{tabular}{| c | c | c | }
\hline 
{\sf {$L$ ($km$)}} & {\sf
{$\rho_{avg}$ ($gm/cm^3$)}}
 & {\sf $E_s^R$ 
(GeV)}  
\\

        \hline
	 295,~732 &   2.3  &  0.20   \\
	 \hline
	  2900  &    3.3   &  0.14  \\
	  \hline
	   7330   &    4.2  & 0.11  \\
	   \hline
	    12000  & 7.6 &  0.06  \\
	    \hline
	    \end{tabular}
	    \caption[]{\footnotesize{Values of  $E_s^R$ 
	    are listed as a function of baseline or the density $\rho$. 
	    See text for details.  
	    The
	    value of $\sin^2 2\theta_{12} = 0.8$ or $\cos 2\theta_{12}=0.43$ is
	    used for evaluating $E_s^R$. }} \label{ettable}
	    \end{center}
	    \end{table}

For energies $E << E_s^R$, the matter term is negligible and we obtain 
the vacuum limit of the survival probability. Its form is similar to
eq.~(\ref{eqpmumu0}) with the vacuum angles and mass-square differences
taking the place of the corresponding matter dependent quantities.
For energies $E >> E_s^R$ or  
$A >> \Delta_{21}$, we get 
\be
\Delta_{21}^m \simeq A - \Delta_{21} \cos 2\theta_{12} 
\ {\rm and} \ \theta_{12}^m \simeq \pi/2.
\label{del21m}
\ee
%
%
Substituting these in eq.~(\ref{eqpmumu0}), we get 
\be
{\rm{P_{\mu \mu}^m}}
=  1 - \sin^2 2\theta_{23}  \sin^2 \left[ \left( \Delta_{31} -
 \Delta_{21} c^2_{12}\right) \frac{L}{4E} \right]. \label{exalp}
\ee
Here we define $c_{ij} = \cos \theta_{ij}$ and $s_{ij} = \sin \theta_{ij}$.
We note that the survival probability in matter is independent of 
the matter term because of cancellations \cite{deGouvea:2005hk}. 
But it must be stressed that 
this probability is in general not the 
same as the vacuum survival probability. It coincides with vacuum $\pmumu$
only in the limit $\Delta_{21}L/4E \ll 1$ and if only terms linear in
the small parameter $\Delta_{21}/\Delta_{31}$ are retained in both 
expressions. In this limit both $\pmumu$ in vacuum and $\pmumum$ 
from eq.~(\ref{exalp})
are given by
\be
P_{\mu\mu} \simeq 1 - \sin^2 2\theta_{23} \left[ \sin^2 \frac{\Delta_{31} L}{4E} - \frac{c_{12}^2 \Delta_{21} L}{4E} \sin \frac{\Delta_{31} L}{2E} \right]
\label{pmumustrong}
\ee
{{
It is to be noted that the lengths involved in the long baseline 
experiments T2K \cite{Itow:2001ee} ($L=295$ Km) and MINOS \cite{Michael:2006rx,MINOS}
and NO$\nu$A \cite{Ayres:2002nm,Ayres:2004js}  ($L = 732/810$ Km)   
are short enough such that
the approximation $\Delta_{21}L/4E \ll 1$ is valid.
The energy of T2K ($E \geq 0.5$ GeV) is a few times $E_s^R$ whereas the 
energies in MINOS and NO$\nu$A 
($E \geq 2$ GeV)
are much larger than $E_s^R$ as can be seen 
from the first row of Table 1. 
Therefore for MINOS
and NO$\nu$A eq.~(\ref{exalp}) is valid to a very good 
approximation. 
It is interesting to consider the precise difference between 
$P_{\mu\mu}^m$ measured by  
T2K \cite{Itow:2001ee}  
and the corresponding vacuum probability in eq.~(\ref{pmumustrong}).
Expanding terms in eq.~(\ref{eqpmumu0})
to first order in the small parameter $\Delta_{21}L/4E$,}} we get
\be
P_{\mu\mu}^m \simeq 1 - \sin^2 2\theta_{23} \left[ \sin^2 \frac{\Delta_{31}^m L}{4E} - \frac{(c_{12}^m)^2 \Delta_{21}^m L}{4E} \sin \frac{\Delta_{31}^m L}{2E} \right]
\label{pmumut2k}
\ee
In this expression, $(c_{12}^m)^2$ is very small $(\sim 0.09)$
because $\theta_{12}^m$ is just a little lower than $\pi/2$.
This multiplies the small term $\Delta_{21}^m L/E$ (which is of the same order as
$\Delta_{21} L/E$). Hence the second term in the square bracket can be neglected
and we have
\bea
P_{\mu\mu}^m &\simeq& 1 - \sin^2 2\theta_{23} \sin^2 \frac{\Delta_{31}^m L}{4E} \nonumber \\
&\simeq& 1 - \sin^2 2\theta_{23} \left[ \sin^2 \frac{\Delta_{31} L}{4E} - \frac{0.55 \Delta_{21} L}{4E} \sin \frac{\Delta_{31} L}{2E} \right] \nonumber \\ 
\label{pmumut2knew}
\eea
Comparing this expression with eq.~(\ref{pmumustrong}) above, 
we find that the only difference is the replacement of the 
factor $c_{12}^2 = 0.69$ by 0.55 in the coefficient of the second term. 
But since this term is suppressed by the small parameter 
$\Delta_{21} L/4E = 0.04$, compared to which the contribution 
of the first term ($\sin^2 \Delta_{31} L/4E$) is about 0.9 for T2K, 
the difference in the coefficient causes a negligible change in the value 
of $P_{\mu\mu}$. The difference in magnitude between $P_{\mu\mu}$ 
from eq.~(\ref{pmumustrong}) and eq.~(\ref{pmumut2knew}) above is less 
than 0.5$\%$. This makes eq.~(\ref{pmumut2knew}) equivalent 
(to a very good approximation) to eq.~(\ref{exalp}) 
in this limit \cite{deGouvea:2005hk}.  

 
\section{Hierarchy Sensitivity} 

It is clear from eq.~(\ref{exalp}) that if the 
magnitude of $(\Delta_{31} - \Delta_{21} c^2_{12})$
is different for NH and IH, 
$\pmumu$ will also be different for the two hierarchies.
Before computing the difference in $\pmumu$,
we need to figure out the difference 
in the magnitude of the quantity $(\Delta_{31} - \Delta_{21} c^2_{12})$
with a change in the hierarchy. We investigate this change
in the light of various assumptions made in the literature.


$\bullet$
First we consider $\Delta_{31}(IH)=-\Delta_{31}(NH)$.
This assumption amounts to identifying $\Delta_{31}$ with $\Delta m_{atm}^2$
determined by Super-K \cite{Ashie:2005ik} and is widely used in the 
literature for situations where the difference between $|\Delta_{31}|$
and $|\Delta_{32}|$ lies well below experimental errors,
and is thus immaterial. 
It leads to
%
\bea
& \: & \mathrm{P}_{\mu \mu}^m({\mathrm {NH}}) - {\rm{P_{\mu \mu}^m}}({\mathrm {IH}})
 =   \nonumber \\
& \; & \sin^2 2\theta_{23}
\left[ \sin^2 \left[(\Delta_{31} + \Delta_{21} c_{12}^2) L/4E \right] \right. \nonumber \\ 
&-& \left. \sin^2 \left[(\Delta_{31} - \Delta_{21} c_{12}^2) L/4E \right] \right].
\label{nh-ih-full}
\eea

Clearly, this assumption is untenable for the situation under consideration here, where the difference between these two mass differences is crucial.   

$\bullet$
The other assumption often used 
is $\Delta_{31}(IH)=-\Delta_{32}(NH)=-\Delta_{31}(NH)+\Delta_{21}$.
This is equivalent to the statement that the largest mass squared difference
has the same magnitude for both NH and IH.
It leads to
\bea
& \: & (\Delta_{31}(IH) - \Delta_{21} c^2_{12})
= (-\Delta_{31}(NH)+\Delta_{21} (1 - c^2_{12})) \nonumber \\
&=& -(\Delta_{31}(NH) - \Delta_{21} s^2_{12}).
\eea
Substituting this in eq.~(\ref{exalp}) gives 
\bea
& \: & \mathrm{P}_{\mu \mu}^m({\mathrm {NH}}) - {\rm{P_{\mu \mu}^m}}({\mathrm {IH}})
 =   \nonumber \\
& \; & \sin^2 2\theta_{23}
\left[\sin^2 \left[(\Delta_{31} - \Delta_{21} s_{12}^2) L/4E \right] \right. \nonumber \\
&-& \left. \sin^2 \left[(\Delta_{31} - \Delta_{21} c_{12}^2) L/4E \right] \right].
\label{nh-ih-full2}
\eea

This amounts to making an ad hoc assumption which currently is unsupported by experimental evidence.


$\bullet$
If we do not make any assumptions but instead consider
the question: what information do the experiments give regarding
the magnitudes of $\Delta_{31}$ and $\Delta_{32}$ for NH and IH ? 
Ongoing experiments such as MINOS and future experiments such as 
T2K and NO$\nu$A measure the muon neutrino survival probability.
In the two flavour limit, $\theta_{13}=0$ and $\Delta_{21}=0$, 
this is given by  
\be
P_{\mu \mu}^{\mathrm 2-fl} = 1 - \sin^2 2 \theta_{23} \sin^2
\left( \frac{\Delta_{31} L}{4E} \right).
\label{pmumu2fl}
\ee
Analyzing the data of these experiments, in this limit, gives us
the magnitude of $\Delta_{31}$. This is the reason behind the 
widely used assumption $\Delta_{31}(IH)=-\Delta_{31}(NH)$, quoted
at the beginning of this section. 

The survival probability 
for T2K and MINOS/NO$\nu$A  
is given by  eq.~(\ref{exalp}) to a very good 
approximation  as shown in the previous section. 
The three flavour effects appear in the form of non-zero $\Delta_{21}$.
We note that eq.~(\ref{exalp}) can be obtained from
eq.~(\ref{pmumu2fl}) by the replacement $\Delta_{31} \to (\Delta_{31}
-c_{12}^2 \Delta_{21})$. Therefore the "atmospheric" mass-squared 
difference measured by MINOS or T2K is not $\Delta_{31}$ (or $\Delta_{32}$)
but the combination 
\cite{deGouvea:2005hk,Nunokawa:2005nx}.
\be
\Delta m^2_{atm} = \Delta_{31} - c_{12}^2 \Delta_{21} = s_{12}^2 
\Delta_{31} + c_{12}^2 \Delta_{32}.
\label{deltatm}
\ee

Thus, in the limit $\theta_{13}=0$, MINOS or T2K measure the magnitude of 
the mass-squared difference $\Delta m^2_{atm}$ defined in 
eq.~(\ref{deltatm}). This quantity is positive for NH and negative
for IH.
Then from eq.~(\ref{deltatm}) one can easily derive  
$\Delta_{31}(IH)=-\Delta_{31}(NH)+2 c_{12}^2 \Delta_{21}$.
With this relation between $\Delta_{31}(NH)$ and $\Delta_{31}(IH)$, 
we see that $\pmumum$ in eq.~(\ref{exalp}) has no hierarchy sensitivity
\cite{deGouvea:2005hk}.
Atmospheric neutrino data with energies $E >> E^R_s$ measure the above $\pmumum$,
and hence contain no hierarchy sensitivity. 
This statement is true independent of the precision
with which $\Delta m^2_{atm}$ can be measured.  
Any deviation from this prediction is likely to come from the regions 
where the approximations
made in obtaining eq.~(\ref{exalp}) are not exact \cite{deGouvea:2005hk,deGouvea:2005mi}. 
This is equivalent to saying that a combination 
of two experiments 
can give hierarchy sensitivity only if the mass-squared differences
(i.e. the frequencies) measured by the two experiments are different. 



In Figure~\ref{fig:hdiff}, the hierarchy difference for $\pmumum$ with $\theta_{13}=0$
is plotted 
as a function of the neutrino energy $E$ for a baseline of 12000 Km using the three definitions of the hierarchy discussed in this section, {\it{i.e.}} (a)  $\Delta_{31}(IH) = -\Delta_{31}(NH)$,
(b) $\Delta_{31}(IH) = -\Delta_{31}(NH) + \Delta_{21}$ and (c)  $\Delta_{31}(IH) = -\Delta_{31}(NH) + 2 c_{12}^2 \Delta_{21}$. Note that the  assumptions a) and b) lead to a large but spurious sensitivity to the hierarchy, as reflected in the large values of the hierarchy difference for $\pmumum$. This difference is greater for case (a) than case (b). 
Case (c) gives the hierarchy difference for $\pmumum$ for the third hierarchy definition,
which is based on the experimental measurement of
$\Delta m_{atm}^2$ by accelerator experiments.
This implies that we need this measurement from experiments such as T2K
or NO$\nu$A to explore the hierarchy sensitivity
in atmospheric neutrino data.
In generating the plots, $\pmumu$ is obtained by the numerical
integration of the neutrino evolution equations.

We see that the hierarchy sensitivity is quite small for case (c). 
A difference $\simeq 0.2$ can be observed in the numerical plots 
at lower energies ($E \simeq 1$ GeV) for this very long baseline of 12000 Km. 
This is because the expressions given in eq.~(\ref{eqpmumu0}) and eq.~(\ref{exalp}),
derived in the constant density approximation, do not hold good at such values of E and L.
But this difference is very small and would be washed out
even with  optimistic values
of energy and angular resolution of the detector.




From Table 1, we see that 
for multi-GeV atmospheric neutrinos the energy E will always be 
larger than $E_s^R$ and so it will be difficult to find a regime where the 
probability will be different than that 
given by  eq.~(\ref{exalp}). Table 1 also shows that
 $E_s^R$ is larger for shorter baselines.
Thus, we expect that
a departure from the form given in eq.~(\ref{exalp}) is most likely 
for the 
baselines of $3000-6000$ Km and neutrino beam
energy of $0.2-0.7$ GeV. These scenarios were analysed in 
\cite{deGouvea:2005mi} for superbeam and neutrino factory experiments. 

If we look at the region in atmospheric neutrino experiments
where the neutrino energy is moderately larger than $E_s^R$ (in the range 0.2 - 0.6 GeV), 
the matter modified mixing angle $\theta_{12}^m$ will be less
than $\pi/2$. Hence, we can't use $\pmumu$ in eq.~(\ref{exalp})
but need to go to the full expression given in eq.~(\ref{eqpmumu0}).
This equation is in principle sensitive to hierarchy due to the matter term
$A$ and the difference between $\Delta_{31}$ and $\Delta_{32}$. 
However, this difference oscillates rapidly with energy
and with baseline. To measure it, one would require extraordinary neutrino energy and angular resolution and
very large statistics. Such angular resolution is impossible
to achieve in the relevant sub-GeV energy range, because
quasi-elastic neutrino scattering, which has a broad distribution in the direction of the final state lepton, 
dominates in this range. Additionally, most atmospheric detectors with target nuclei have
an inherent limit on the energy resolution possible, set by Fermi motion 
of the bound nucleons. Such energies are typically $\sim 100$ MeV, 
and thus are a significant fraction of the energy of the final state 
lepton in the range of interest here.

In Figure 2, we plot the difference in $\pmumum$ for NH and IH as a function of baseline for the two energies, 0.5 GeV and 1.0 GeV.
These figures show a moderate to substantial difference for the case of no smearing. However, this difference becomes less than 0.02 
when we include an angular smearing with $\sigma_{\theta}=5^{\circ}$. 
This value of $\sigma_{\theta}$ is much smaller than the width 
of the angular distribution for quasi-elastic events. Note that no energy 
smearing was included in generating this figure. Thus we see that even with an ideal energy resolution, the difference
in $\pmumum$ for the two hierarchies becomes tiny. If a moderate energy smearing is included the difference is likely to vanish. Hence it is not possible to determine the neutrino mass hierarchy 
using atmospheric neutrinos as a source, no matter how good the detector is.

To summarize, the general requirements for 
hierarchy determination for $\theta_{13} =0$ are 
\begin{enumerate}

\item  $|\Delta m^2_{\mathrm atm}|$ should be measured to a
precision better than 
2\%. 

\item The neutrino energy $E$ should be large enough to
produce a muon but should not be too large compared
to $E_s^R$.

\item The baseline $L$ should be such that $\Delta_{21} L/E \gtrsim 1$.

\item Excellent 
energy  and angular resolution are necessary,  so that rapid oscillations in 
$P_{\mu \mu}$ 
can be resolved. 

\end{enumerate} 

Atmospheric neutrinos clearly do not satisfy the last criterion.  In 
a beam experiment, however, $\sigma_{\theta}$ is zero, so the problem of angular resolution is automatically resolved. However, it must still meet the requirement of superior  energy resolution, good  enough to resolve the closely spaced peaks shown in
case (c) of Figure 1.
It is apparent from the size of the peaks that the experiment needs to 
have the capability to gather high statistics \cite{deGouvea:2005mi}. 

\section{Conclusions}

In conclusion, we have studied the feasibility of  determining the
sign of $\Delta_{31}$ for
$\theta_{13} =0$ using atmospheric neutrinos as the source. 
When $\theta_{13}=0$, matter effects induced by it are absent. 
Moreover, $\nu_e$ is decoupled from $\nu_3$,
leading to no  hierarchy dependence in $\pee$ or $\pmue$. This makes 
$\pmumu$ the most suitable remaining channel
for hierarchy determination. 
In this limit,   $\pmumu$ depends on  $\Delta_{21}, \Delta_{31}$ and
$\Delta_{32}$. For NH (IH), $\Delta_{31}$ ($\Delta_{32}$) is the 
highest frequency.
If these two close frequencies  can be resolved along
with their magnitudes, then the hierarchy can be determined.
This requirement  is, in practice, difficult to achieve
by measurement at a single $L/E$. 
Hence it translates to having
two experiments,
one of which measures a single frequency  ($|\Delta m^2_{\mathrm atm}|$), 
which is a known combination of
the
two independent
frequencies ($\Delta_{21}$ and $\Delta_{31}$),  
to high precision. The other experiment then needs to satisfy the 
requirement  $\Delta_{21} L/E \gtrsim 1$ and 
measure a {\it different} combination of these frequencies,
also with high precision. 
The precision requirements imply that the energy and angular resolutions have to be exceptionally
good.  
The $\Delta_{21} L/E \gtrsim 1$ 
requirement implies long baselines. 
The condition for measuring a different frequency enforces 
tapping into $\Delta_{21}$ driven matter effects  which in turn
needs energies 
which are not too far removed from  the solar resonance energy $E_s^R$.

Atmospheric neutrinos offer a broad range of $L/E$ values 
including $\Delta_{21} L/E \gtrsim 1$.
But the precision requirements preclude
any chance of hierarchy determination 
by a single atmospheric experiment on its own. 
We have explored if hierarchy determination is possible 
using atmospheric neutrinos in conjunction
with upcoming accelerator experiments 
such as T2K and NO$\nu$A. For these experiments,
$L/E$ values are such that $\Delta_{31} L/E \sim 1$ and hence $\Delta_{21} L/E << 1$.
In this approximation, $\pmumum$
depends on the single effective frequency $\Delta_{31} -c_{12}^2  \Delta_{21}$. {{For atmospheric neutrinos with energies above 1 GeV and satisfying the condition  $\Delta_{21} L/E \gtrsim 1$, one will
measure this same frequency. Thus, even if one had a high precision atmospheric neutrino experiment detecting multi-GeV neutrinos, this combination 
would not satisfy the  requirement of different frequency measurement discussed above.

If, on the other hand, one had an atmospheric experiment capable of
measuring lower energies (which would have the consequence of widening the range of $L$ required 
to satisfy $\Delta_{21} L/E \gtrsim 1$ and allowing the inclusion of shorter baselines), it is likely that
the different frequency condition could be met. However, the
precision requirements on energy and angular resolution become very
demanding in this case. 
In particular, the energy resolution has to be better than the typical Fermi energy of a bound nucleon.
Also, given that quasi-elastic scattering provides the dominant contribution to the event rates for sub-GeV neutrinos, the precision requirement on angular resolution is impossible to realize. Hence neutrino hierarchy determination using atmospheric neutrinos, even in conjunction with another high precision beam experiment, is not possible.


\begin{figure*}[htb]
\centerline
{
\epsfxsize=10.0cm\epsfysize=10.0cm\epsfbox{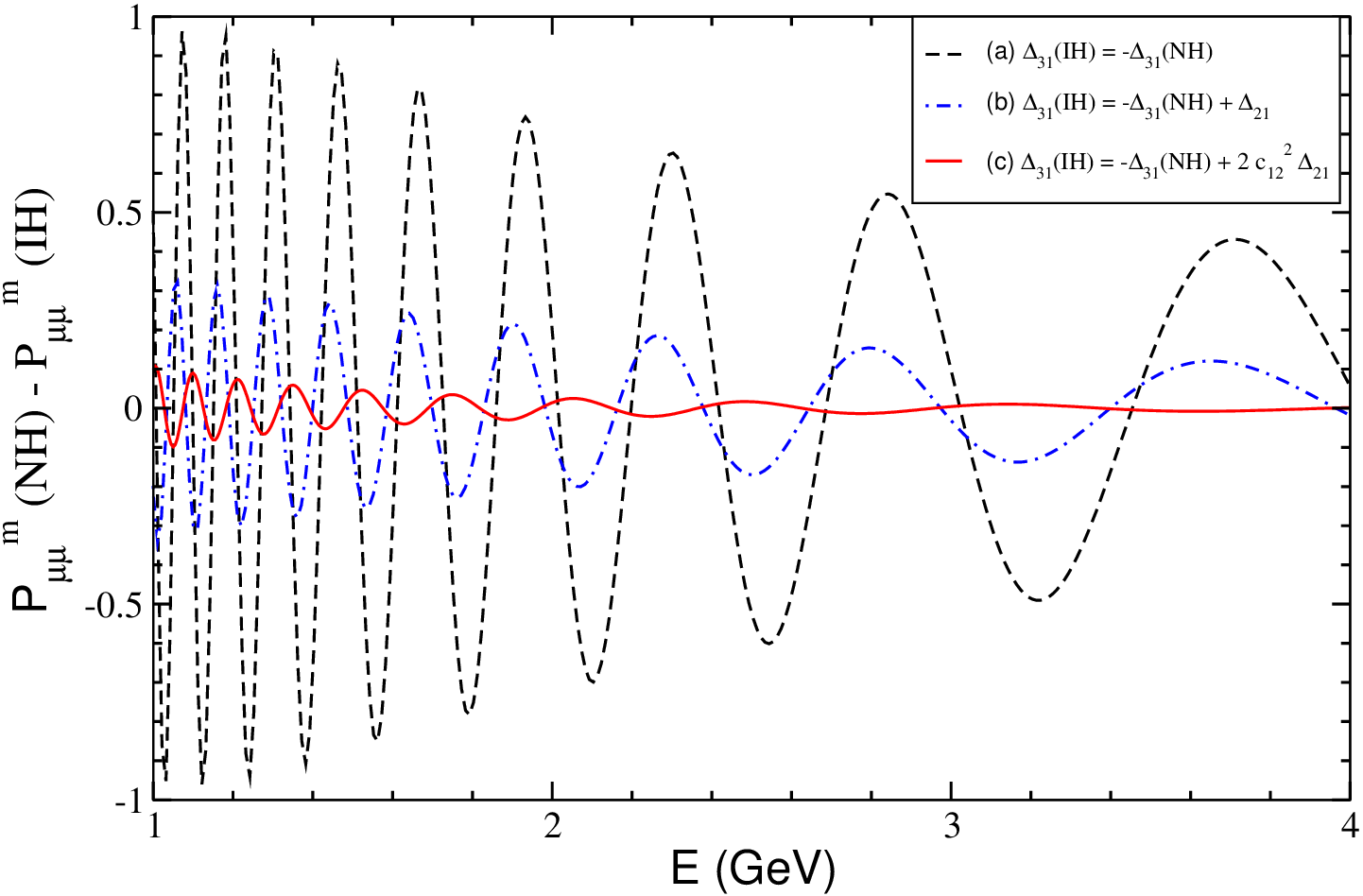}
}
\caption
{Hierarchy difference for $\pmumum$
(exact numerical muon survival probability in matter)
as a function of the neutrino energy E with L = 12000 Km and
$\theta_{13}=0$ for the three assumptions relating $\Delta_{31}(NH)$ and $\Delta_{31}(IH)$.}
\label{fig:hdiff}
\end{figure*}

\begin{figure*}[htb]
\centerline
{
\epsfxsize=9.0cm\epsfysize=9.0cm\epsfbox{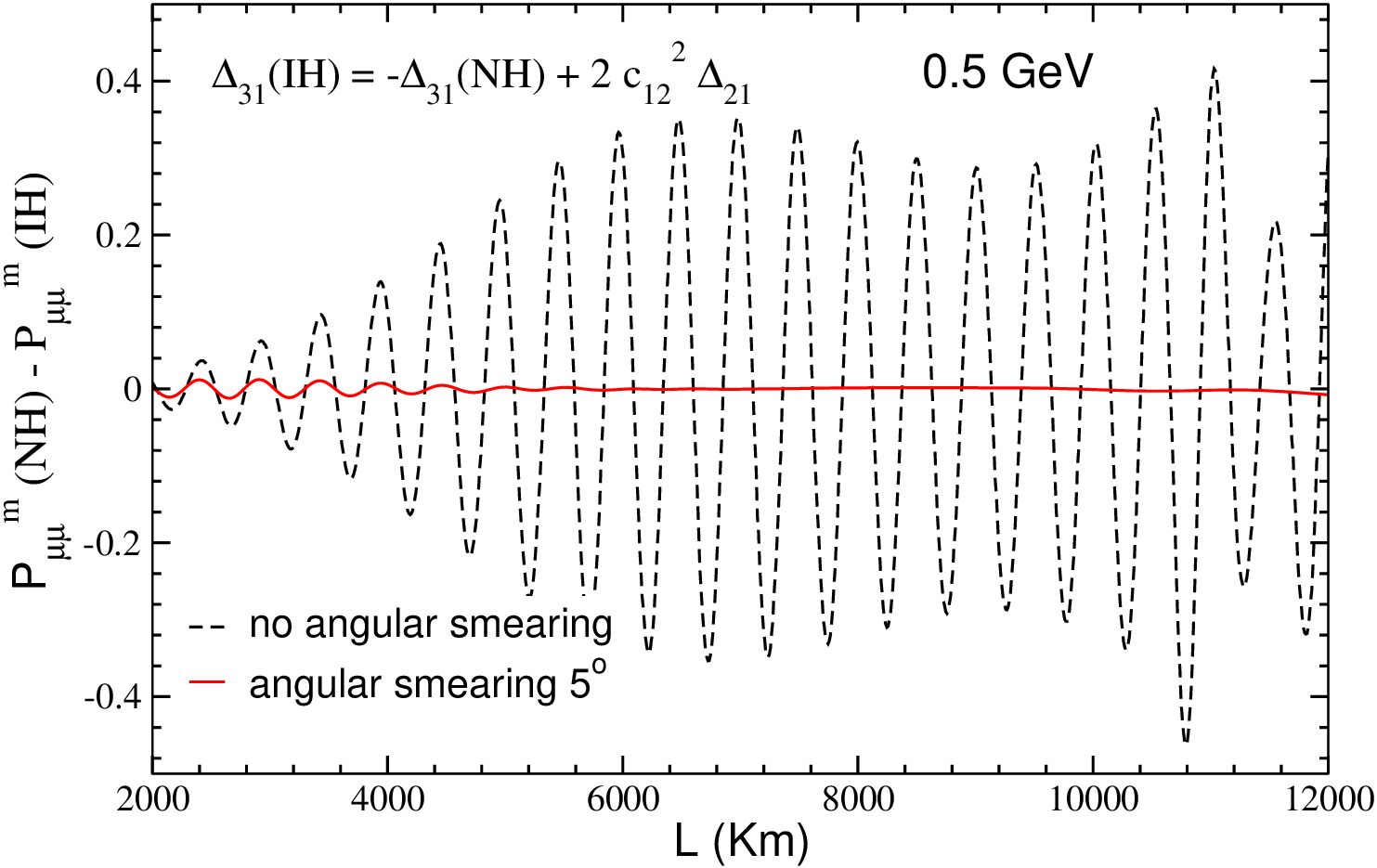}
        \hspace*{0.5ex}
\epsfxsize=9.0cm\epsfysize=9.0cm
                     \epsfbox{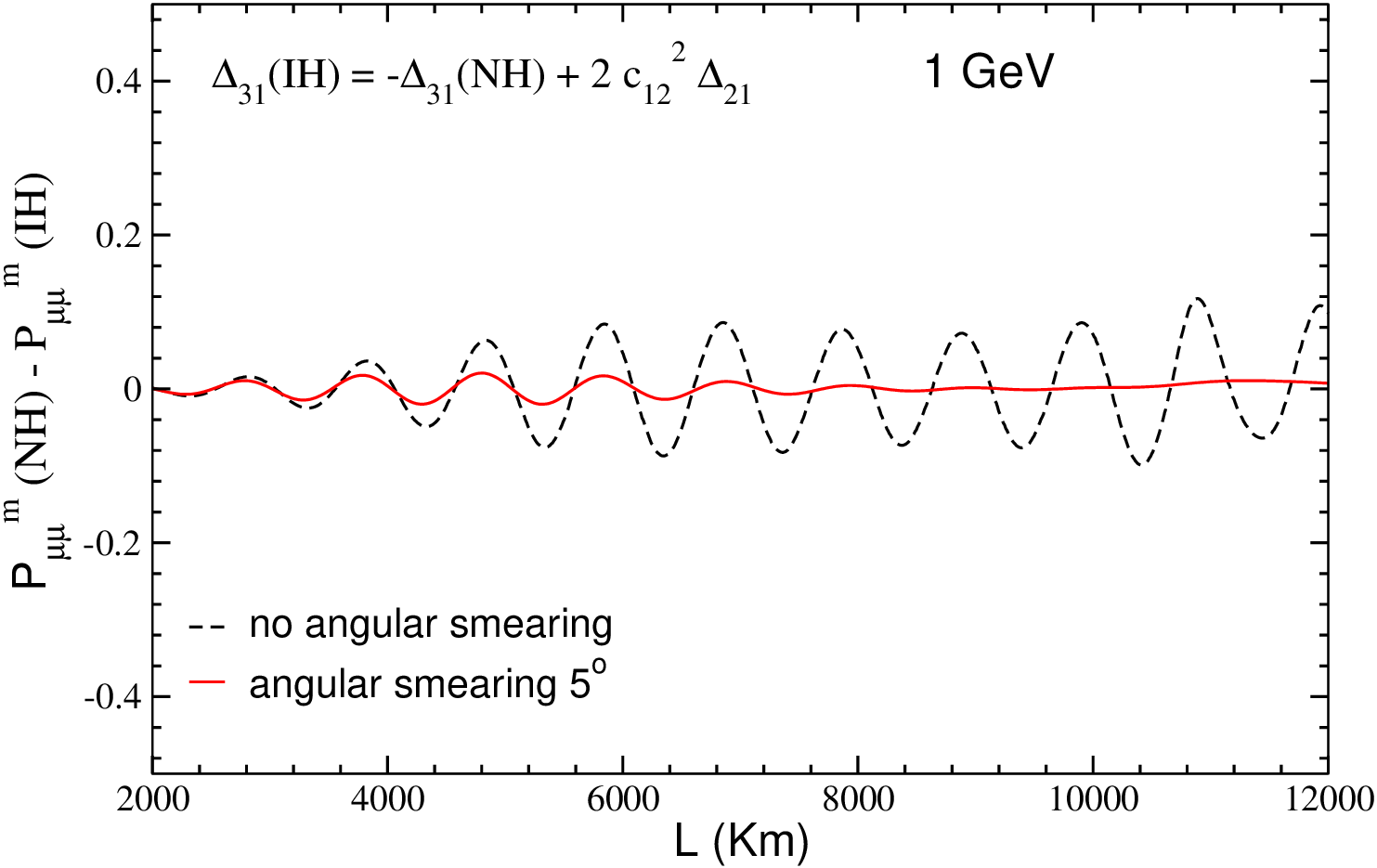}
}
\hskip 5cm
{\bf (a)}
\hskip 7cm
{\bf (b)}
\caption
{Hierarchy difference for $\pmumum$
(exact numerical muon survival probability in matter)
as a function of the baseline L
for
$\theta_{13}=0$ with the T2K definition of $\Delta_{31}(IH)$, for (a) E $=$ 0.5 GeV and (b) E $=$ 1 GeV. 
The difference without angular smearing and with an angular smearing of 5$^{\circ}$ are shown.
}
\label{fig:hdiff2}
\end{figure*}

%
%
%
%
%
%

\section{Acknowledgements} 
The authors thank Sandip Pakvasa for pointing out the importance
of defining the relation between mass-squared differences for different hierarchies 
in terms of experimentally measured quantities. 
They would also like to thank Thomas Schwetz, Walter Winter, Michele Maltoni,
Patrick Huber, Toshihiko Ota and Ravishanker Singh
for discussions and  would like to acknowledge 
the involvement of Poonam Mehta at an initial stage of this work. 
R.G. and S.G. acknowledge support from the XIth plan
neutrino project of Harish-Chandra Research Institute. S.G. and
S.U.S acknowledge partial support from a BRNS project . R.G.
acknowledges the  hospitality of the CERN Theory Division and the University
of Wisconsin-Madison phenomenology group and S.G.
acknowledges hospitality of the TIFR theory group during the
finishing stage of this work. P.G. acknowledges partial support by a Max Planck
India Partnergroup grant in TIFR, Mumbai.






\bibliographystyle{apsrevwinter}

\bibliography{myrefjan_08}

\end{document}